\documentclass[10pt]{amsart}

\usepackage{color}
\usepackage{amsmath}
\usepackage{amssymb}
\usepackage{float}
\usepackage{graphicx}
\newtheorem{Teorema}{Theorem}[section]
\newtheorem{proposicion}{Proposition}[section]

\newcommand{\RR}{{\mathbb R}}

\newsavebox{\savepar}

\def\nn{\mathbb{N}}

\def\G{\mathcal{G}}

\def\P{\mathcal{P}}
\def\Q{\mathcal{Q}}

\def\rI{\rm I}

\def\epsilon{\varepsilon}
\def\leq{\leqslant}
\def\geq{\geqslant}

\def\x{{\bf x}}
\def\y{{\bf y}}
\def\z{{\bf z}}

\def\pp{\mathbb{P}}
\def\ee{\mathbb{E}}

\def\ms{{\medskip\noindent}}
\def\bs{{\bigskip\noindent}}

\def\bG{\bar{\G}}
\def\bV{\bar{V}}
\def\bU{\bar{U}}
\def\bA{\bar{A}}

\def\Id{{\rm Id}}
\def\Od{{\rm Od}}

\def\d{\partial}

\begin{document}
\title{Dominant Vertices in Regulatory Networks Dynamics}
\author{Beatriz Luna \& Edgardo Ugalde}

\maketitle

\begin{center}
\begin{minipage}{13truecm}
\begin{small}
Instituto de F\'\i sica, Universidad Aut\'onoma de San Luis Potos\'\i , \\
Av. Manuel Nava 6, Zona Universitaria, 78290 San Luis Potos\'\i , M\'exico.
\end{small}
\end{minipage}
\end{center}

\begin{abstract}
Discrete--time regulatory networks are dynamical systems
on directed graphs, with a structure inspired on natural systems of 
interacting units. There is a natural notion of determination 
amongst vertices, which we use to classify the nodes of the network,
and to determine what we call {\it sets of dominant vertices}.
In this paper we prove that in the asymptotic regime, the 
projection of the dynamics on a dominant set allows us to determine the 
state of the whole system at all times.
We provide an algorithm to find sets of dominant vertices, and we test its
accuracy on three families of theoretical examples. Then, by using the
same algorithm, we study the relation between the structure of the underlying 
network and the corresponding dominant set of vertices. 
We also present a result concerning the inheritability of the dominance
between strongly connected networks.

\bs {\it Keywords}: 
Regulatory Dynamics on Networks, Dominant Vertices, Controlled Trajectories.

\ms {\it MSC:} 37L60, 37N25, 05C69.

\end{abstract}


\section{Introduction}

\ms In this work we consider a class of discrete--time dynamical systems on networks,  
inspired on natural systems of interacting units as for instance the gene regulatory 
networks. This class of discrete--time models was first introduced in~\cite{VL05}, 
and in~\cite{CFML06} their low dimensional representatives were studied in full detail. 
These systems can be thought as a discrete--time alternative to systems of 
piecewise affine differential equations extensively studied in~\cite{dJ02, E00, MPO95}, 
and to finite state models, better known as logical networks, 
which have been used in~\cite{K69, GK74, TT95} for instance. In each of these modeling 
strategies, the interacting units have a regular behavior when taken separately, 
but are capable to generate global complex dynamics when arranged in a complex interaction 
architecture. In the discrete--time dynamical systems considered here, all interacting 
units evolve synchronously at discrete time steps, and the level of activity of each one 
of them varies according a rule involving its neighboring units in the network.

\ms The interaction architecture may allows to reconstruct the 
dynamics of the whole systems from the observation of a small number of units. 
The record of the activity of those units during the evolution  
contains all the information needed to determine the state of the system at 
all times. The aim of the present work is to give a topological characterization 
of these determining units, in the framework of the discrete--time dynamical 
systems on networks mentioned above. This characterization is achieved  
via a classification of the vertices of the network, according to a determination 
relation we define below. The main result states that the vertices in the 
leading class capture the whole dynamics of the network, i.~e., if two orbits
are such that their restriction to the leading vertices approach to each other, 
then they asymptotically coincide at each vertex of the network. On one hand, 
the vertices in the leading class, or dominant vertices as we call them, 
are sites to record data in order to reconstruct the state of the whole 
system at each time. On the other hand, those vertices can be used as controlling
sites. If one forces the configuration of the system to follow a particular
trajectory at those sites, then this forcing will spread through the whole network.

\ms The paper is organized as follows. After some preliminary definitions and notations
given in the next section, we present in Section~\ref{section-dominance} the main 
results, as well as an algorithm determining a set of dominant vertices.
In order to get an insight in the relation between the underlying network and the structure
of the dominant set of vertices, in Section~\ref{section-thexamples} we study 
several families of theoretical examples. Then, in Section~\ref{section-nuexamples},
we perform a numerical study of the dominant set of vertices in Erdo\"s--R\'enyi 
and Barab\'asi--Albert random networks. We finish with some concluding remarks and
general comments.

\bs \section{Preliminaries}~\label{section-preliminaires}\

\ms A discrete--time regulatory network is a 
discrete--time dynamical dynamical system on a network. 
We start with a network $\G=(V,A)$, where $V$ are the vertices and represent 
interacting units, and the arrows $A\subseteq V \times V$ denote interaction between 
them. For each $v\in V$, let $I(v):=\{u\in V:\ (u,v)\in A\}$ be the set of vertices 
influencing $v$, and $O(v):=\{u\in V:\ (v,u)\in A\}$ the set of vertices 
influenced by $v$. To each interaction $(u,v)\in A$, we associate a partition 
$\P_{uv}$ of $[0,1]$,
by a finite number of intervals. Each atom in $\P_{uv}$ corresponds to an activation 
mode for the interaction $(u,v)\in A$, so that each atom in 
$\Q_v:=\prod_{I(v)}\P_{uv}$ is associate to a total activation state 
on the vertex $v\in V$. 
For each unit $u\in V$, the activation partitions $\{\P_{uv}: u\in I(v)\}$ define a 
refinement $\P_u:=\bigvee_{v\in V}\P_{uv}$, composed by the 
intersections $\bigcap_{v\in V}\rI_{uv}$ with $\rI_{uv}\in \P_{uv}$. 
Finally, these refinements give place to the $\#V$--dimensional partition 
$\P:=\prod_v\P_v $. We will denote $\rI(\x)$ to the atom of $\P$ containing $\x$. 

\ms The collective activation mode of the network remains constant inside each atom 
of the partition $\P$ defined above, and the evolution of the system is given 
by the iteration of the piecewise affine transformation 
$F:[0,1]^V\rightarrow [0,1]^V$, such that 
\begin{equation}~\label{din}
\x^{t+1}_v:=F(\x^t)=a \x^{t}_v+(1-a) D(\x^t),
\end{equation}
where $a \in [0,1)$ is the contraction rate related to speed of degradation of the 
activity of the units in absence of interaction, and $D:[0,1]^V\to [0,1]^V$ is a
piecewise constant function representing the regulatory interactions.
The piecewise constant term $D:[0,1]^V\to [0,1]^V$ depends on the interaction 
architecture, and for each $v\in V$ is such that 
\begin{equation}~\label{constant}
D(\x)_v=D_v(\x_u: \ u\in I(v))\equiv D_v\left(\x_{I(v)}\right),  
\end{equation} 
where $D_v:[0,1]^{I(v)}\to [0,1]$, is constant in the atoms of 
$\Q_v$~\footnote{In previous works~\cite{CFML06, LU06} it has been considered less general 
interaction functions defined as follows. For each interaction $(u,v)\in A$ 
there are associated a sign $\sigma_{uv}\in \{-1,1\}$, and a threshold $T_{uv} \in (0,1)$. Only two activation modes are considered: activation if $\sigma_{uv}=1$, and an 
inhibition in the case $\sigma_{uv}=-1$. The interaction function  
$D:[0,1]^V\to [0,1]^V$, is then given by
$D(\x)_v=(\#I(v))^{-1}\sum_{u \in V} H(\sigma_{uv}(\x_u-T_{uv}))$.
with $H:\RR\rightarrow \RR$ the Heaviside function.}. 
Finally, the {\it discrete--time regulatory network} defined by the
interaction function $D$ and the contraction rate $a$, is the discrete--time 
dynamical system $([0,1]^V,F)$, with $F$ as defined in~\eqref{din}. The
interaction architecture, codified in the digraph $(V,A)$, is implicit in
the definition of the interaction function. 

\ms As usual, {\it the orbit corresponding to the initial condition $\x^{0}\in[0,1]^V$}, 
is the sequence $\{\x^t\}_{t\in\nn}$ in $[0,1]^V$ such that $\x^{t+1}=F(\x^t)$ 
for each $t\in\nn$. We will also consider controlled trajectories, which we
define as follows. For $U\subseteq V$ fixed, a {\it $U$--controlled trajectory} is 
any sequence $\left\{\x^{t}\right\}_{t\in\nn}$ in $[0,1]^V$ such that 
$\x^{t+1}_{V\setminus U}=F(\x^{t+1})_{V\setminus U}$. Here, the projection 
$\left\{\x^{t}_U\right\}_{t\in \nn}$, which we also call {\it control term}, is taken arbitrarily.

\ms The discontinuity set $\Delta_{\P}$ of the transformation $F$ is composed by the 
borders of all the rectangles in $\P$. It is a union of $(\#V-1)$--dimensional 
hyperplanes perpendicular to the principal axes of the hypercube $[0,1]^V$. 
We will say that the orbit or controlled trajectory $\{\x^t\}_{t=0}^\infty$ is 
{\it uniformly separated from the discontinuity set}, if there exists $\epsilon >0$ 
such that $\inf_{t\geq 0}\left(\inf_{\z\in \Delta_{\P}}|\x^t-\z|\right)> \epsilon$. 
Following the arguments developed in~\cite{CMU07}, it can be shown that uniform
separation from the discontinuity set is a typical situation. For small enough 
contraction rate, it is should be generic (in the topological sense), with respect
to the choice of the partitions. In general, by using an appropriate distribution
in the set of partitions, the uniform separation could be proved to hold with 
probability one.

\bs \section{Dominance}~\label{section-dominance} \

\ms For vertex sets $U, U' \subseteq V$, we say that {\it $U$ determines $U'$}, 
if $U\supseteq \bigcup_{v\in U'}I(v)$, i.~e., all the arrows with head in $U'$ have
their tails in $U$. We will denoted by $\d U$ the maximal set determined by $U$, 
which is given by
\begin{equation}~\label{determines}
\d U:=\{v\in V:\  I(v)\neq\emptyset\,\text{ and }\,I(v)\subseteq U\}.
\end{equation}

\ms Whenever $U$ determines $U'$, $D(\x)_v$ at the coordinates $v\in U'$ 
can be computed from the values $\x_u$ at coordinates $u\in U$. Indeed, 
if $\x_{U}$ and $\y_{U}$ belong to the 
same rectangle in $\prod_{u\in U}\P_u$, then $D(\x)_{U'}=D (\tilde {\x})_{U'}$. 
Since $D_{U'}\equiv\prod_{u\in U'}D_u\left(\x_{I(u)}\right)$, and
$U\supseteq\bigcup_{v\in U'}I(v)$, then we will use the notation 
$D(\x)_{U'}=D(\x_U)$ in this case.

\ms {\it A vertex set $U\subseteq V$ will be called dominant} if there exists a 
nested sequences of vertex sets,
$U\equiv U_1\subset U_2\subset \cdots U_{d}\subset U_{d+1}\equiv V$, such that
\[\d U_{i}= U_{i+1}\setminus U_i, \ i=1,\ldots,d.\] 
Clearly $U$ uniquely determines the sequence $U_2\ldots U_{d+1}$. We will refer
to its length $d$, as {\it the depth of the dominant set $U$}.
The classical definition of dominance for directed graphs (as given in~\cite{HHS98} 
for instance), stipulates that $U$ is dominant if $V=U\cup \d U$. 
Here we slightly extend this notion to allow determinations in more that one step.

\bs According to our definition, a dominant set of vertices $U\subseteq V$ 
governs the dynamics of the whole network through a chain 
$U=U_1\subset U_2\cdots U_{d+1}=V$ of set, so that for each $i=1,\ldots,d$, $U_i$ 
determines $V_{i}:=U_{i+1}\setminus U_i$. 
Since $\bigcup_{v\in V_i}I(v)\subseteq U_i$, then $D(\x)_v$ at the coordinates 
$v\in V_i$ can be computed from the values $\x_u$ at coordinates $u\in U_i$.
This allows us to establish the following chain of equations,
\begin{eqnarray}~\label{equation-nested}
F (\x_{V_1})&=&a \x_{V_1}+(1-a)D(\x_{U})\\
F (\x_{V_2})&=&a \x_{V_2}+(1-a)D(\x_{U_2}) \nonumber\\ 
            &\vdots & \nonumber\\
F (\x_{V_d})&=&a \x_{V_d}+(1-a)D(\x_{U_d})\nonumber\\
F (\x_{U})&=&a \x_{U}+(1-a)D(\x_{U_{d+1}}),\nonumber
\end{eqnarray}
for all $\x\in [0,1]^V$.

\ms The equations in~\eqref{equation-nested} are the key ingredient behind the proof
of the following theorem. Other important element is the
concept of uniform separation which we remind now. The orbit or 
controlled trajectory $\{\x^t\}_{t=0}^\infty$ 
is uniformly separated from the discontinuity set, if there exists $\epsilon >0$ such that 
$\inf_{t\geq 0}\left(\inf_{\z\in \Delta_{\P}}|\x^t-\z|\right)> \epsilon$. 
We have the following.

\bs
\begin{Teorema}~\label{Teorema1}
Let $U\subseteq V$ be a dominant set of vertices, and 
$\{\x^{t}\}_{t\in\nn},\ \{\y^{t}\}_{t\in\nn}$
two orbits or $U$--controlled trajectories, both uniformly separated from 
the discontinuity set. If  $\lim_{t\to\infty}\left|\x^t_U-\y^t_U\right|=0$, 
then 
$\lim_{t\to \infty}|\x^t-\y^t|=0$.
\end{Teorema}

\ms
\begin{proof} 
Let $U=U_1\subset U_2 \subset U_{d+1}=V$ be the nested sequence associated to $U$,
and for each $i=1,2,\ldots,p$, let $V_i=U_{i+1}\setminus U_i$.
Let $\epsilon >0$ be such that  
$\inf_{t\geq 0}\left(\inf_{\z\in \Delta_{\P}}|\x^t-\z|\right)> \epsilon$ 
and $\inf_{t\geq 0}\left(\inf_{\z\in \Delta_{\P}}|\y^t-\z|\right)> \epsilon$,
and take an arbitrary $\delta < \epsilon/(2\sqrt{d+1})$. 

\ms Starting from time $t=t_0$, we iterate the equations in~\eqref{equation-nested} in 
order to obtain, for each $\tau_0\in\nn$,
\begin{eqnarray}~\label{equation-nested2}
\x^{\tau+t_0}_{V_1}
&=&a^{\tau} \x^{t_0}_{V_1}+(1-a)\sum_{t=0}^{\tau-1}\,a^{\tau-1-t}\,D(\x^{t+t_0}_{U})\\
\x^{\tau+\tau_0+t_0}_{V_2}&=&a^{\tau} \x^{\tau_0+t_0}_{V_2}+(1-a)
           \sum_{t=0}^{\tau-1}\,a^{\tau-1-t}\,D(\x^{t+\tau_0+t_0}_{U_2}) \nonumber\\
&\vdots& \nonumber\\
\x^{\tau+(d-1)\tau_0+t_0}_{V_d}&=&a^{\tau} \x^{(d-1)\tau_0+t_0}_{V_d}+(1-a)
                      \sum_{t=0}^{\tau-1}\,a^{\tau-1-t}\,D(\x^{t+(d-1)\tau_0+t_0}_{U_{d}}),\nonumber
\end{eqnarray}
and all $\tau\geq 0$. A similar set of equations is satisfied by $\y^t$.
Since $\lim_{t\to\infty}\left|\x^t_U-\y^t_U\right|=0$, and 
both are uniformly separated from the discontinuity set, then there exists $t_0\in\nn$ 
such that for each $t\geq t_0$,
$\x_U^t$ and $\y_U^t$ belong to the same atom in $\prod_{u\in U}\P_u$, and
such that $\left|\x_U^t-\y_U^t\right|<\delta$.
Hence we have $D(\x^t_U)=D(\y^t_U)$ for $t\geq t_0$, and taking into account
the first equation in~\eqref{equation-nested2} we obtain
\begin{eqnarray*}
\left|\x_{U_2}^{\tau+t_0}-\y_{U_2}^{\tau+t_0}\right|
&=&
\left|
\left(\x_{U}^{\tau+t_0}-\y_U^{\tau+t_0}\right)\oplus \left(a^{\tau} 
\left(\x^{t_0}_{V_1}-\y_{V_1}^{t_0}\right)+(1-a)
\sum_{t=0}^{\tau-1}\,a^{\tau-1-t}\,\left(D(\x^{t+t_0}_{U})-D(\y^{t+t_0}_{U})\right)\right)
\right|\\
&=&\left|
\left(\x_{U}^{\tau+t_0}-\y_U^{\tau+t_0}\right)\oplus
a^{\tau} \left(\x^{t_0}_{V_1}-\y_{V_1}^{t_0}\right)
\right|\\
&=&\sqrt{\left|\x_{U}^{\tau+t_0}-\y_U^{\tau+t_0}\right|^2 +
a^{2\tau}\ \left|\x^{t_0}_{V_1}-\y_{V_1}^{t_0}\right|^2}.
\end{eqnarray*}
By taking $\tau_0$ such that $a^{\tau_0}< \delta$, we ensure that
$\left|\x_{U_2}^t-\y_{U_2}^t\right|< \sqrt{2}\,\delta < \epsilon/2$ for each 
$t\geq \tau_0+t_0$. Therefore $\x^t_{U_2}$ and $\y_{U_2}^t$ belong to the same 
atom in $\prod_{u\in U_2}\P_u$ for all $t\geq \tau_0+t_0$, which implies
$D(\x^{\tau}_{U_2})=D(\y^{\tau}_{U_2})$ for all $\tau\geq\tau_0+t_0$. 
By using this, and the second equation 
in~\eqref{equation-nested2}, we get
\begin{eqnarray*}
\left|\x_{U_3}^{\tau+\tau_0+t_0}-\y_{U_3}^{\tau+\tau_0+t_0}\right|
&=&\left| \left(\x_{U_2}^{\tau+\tau_0+t_0}-\y_{U_2}^{\tau+\tau_0+t_0}\right)
\oplus
 a^{\tau} \left(\x^{\tau_0+t_0}_{V_2}-\y_{V_2}^{\tau_0+t_0}\right)
\right|\\
&=&\sqrt{\left|\x_{U_2}^{\tau+\tau_0+t_0}-\y_{U_2}^{\tau+\tau_0+t_0}\right|^2+
 a^{2\tau}\left|\x^{\tau_0+t_0}_{V_2}-\y_{V_2}^{\tau_0+t_0}\right|^2}.
\end{eqnarray*}
Therefore, $\left|\x_{U_3}^{t}-\y_{U_3}^{t}\right|<\sqrt{3}\,\delta <\epsilon/2$ for each 
$t\geq 2\ \tau_0+t_0$, which ensures that $D(\x^t_{U_3})=D(\y^t_{U_3})$ 
for each $t\geq 2\ \tau_0+t_0$.
Iterate this argument, we finally obtain
\begin{eqnarray*}
\left|\x^{\tau+(d-1)\tau_0+t_0}-\y^{\tau+(d-1)\tau_0+t_0}\right|
&\equiv& 
\left|\x_{U_{d+1}}^{\tau+(d-1)\tau_0+t_0}-\y_{U_{d+1}}^{\tau+(d-1)\tau_0+t_0}\right|\\
&=&\left|\left(\x_{U_d}^{\tau+(d-1)\tau_0+t_0}-\y_{U_d}^{\tau+(d-1)\tau_0+t_0}\right)
\oplus a^{\tau} 
\left(\x^{(d-1)\tau_0+t_0}_{V_d}-\y_{V_d}^{(d-1)\tau_0+t_0}\right)
\right|\\
&=&\sqrt{\left|\x_{U_d}^{\tau+(d-1)\tau_0+t_0}-\y_{U_d}^{\tau+(d-1)\tau_0+t_0}\right|^2
+a^{2\tau}\left|\x^{(d-1)\tau_0+t_0}_{V_d}-\y_{V_d}^{(d-1)\tau_0+t_0}\right|^2}.
\end{eqnarray*}
In this way we have prove that 
$\left|\x^t-\y^t\right|\leq \sqrt{d+1}\,\delta$, for all $t \geq d\tau_0+t_0$, and
since $\delta>0$ is arbitrary, the theorem follows.  
\end{proof}

\ms According to this theorem, the dominant set of vertices 
$U\subseteq V$ captures the dynamics of the whole network $\G:=(V,A)$ through the chain 
$U=U_1\subset U_2\cdots U_{d+1}=V$, i.~e., in the asymptotic regime,
the projection of the dynamics on the dominant set allows us to determine the 
state of the whole system. This is achieved with a delay which depends 
on the depth of the dominant set, the contraction rate, and on the particular 
trajectory. If state of the dominant set is forced to follow a prescribed trajectory, 
the asymptotic dynamics of the whole system will depend only on this
forcing. In this case we say that the dynamics of the network have been controlled.

\ms Dominance is inherited from a completely connected networks to their completely 
connected subnetworks. To be more precise in this respect, let us state the following.

\bs 
\begin{proposicion}[Inhering Dominance]~\label{proposicion1}
Let $\G=(V,A)$ be a strongly connected network, and $\bG:=(V,\bA)\preccurlyeq \G$, a 
subnetwork strongly connected as well. Let $U\subseteq V$ be a dominant set of vertices 
for $\G$. Then, there exists a set of vertices $\bU\subseteq U$, which is dominant for $\bG$.
\end{proposicion}

\ms \begin{proof} 
For $v\in V$, let $I_{\bG}(v):=\{u\in V:\ (u,v)\in \bA\}$ and 
$I_{\G}(v):=\{u\in V:\ (u,v)\in A\}$. Then, for $\bV\subseteq V$,
let 
$\d_{\bG}\bV:=\{v\in V:\ I_{\bG}(v)\neq\emptyset\,\text{ and }\, I_{\bG}(v)\subseteq \bV\}$,
and similarly for $\d_{\G}V'$.

\ms Since $\bA\subseteq A$, and $\bG$ is strongly connected, then, for each $\bV\subseteq V$,
we have $\d_{\bG}\bV\supseteq \d_{\G}\bV$. 
Indeed, $v\in \d_{\G}\bV$ if and only if $I_{\G}(v)\neq\emptyset$ and 
$I_{\G}(v)\subseteq \bV$. 
Since $I_{\bG}(v)\subseteq I_{\G}(v)$ and $\bG$ is strongly connected, then 
$I_{\bG}(v)\neq\emptyset$ and $I_{\bG}(v)\subseteq \bV$ as well. 
Therefore $v\in\d_{\bG}\bV$.

\ms Let $U=U_1\subset U_2\cdots \subset U_{d+1}=V$ be the nested sequence associated to $U$.
For each $i=1,2,\ldots,d$, we have $U_{i+1}=U_i\cup\d_{\G}U_i$. Hence, by taking
$\bU=\bU_1:=U$, and for each $i=1,2,\ldots,p$, $\bU_{i+1}:=\bU_i\cup\d_{\bG}\bU_i$,
we ensure that $\bU_i\supseteq U_i$ for each $i=1,2,\ldots,p$. 
In this way we associate
to $\bU=U$ a nested sequence $\bU=\bU_1\subset \bU_2\cdots \subset \bU_{\bar{d}+1}=V$,
with $\bar{d}:=\min\{1\leq i\leq d:\  \bU_{i+1}=V\}$,
which satisfies the requirements for $\bU$ to be a dominant set. 
\end{proof}

\bs\subsection{Determining a Dominant Set}\

\ms The following is an algorithm we have implemented for the determination of a
dominant set of vertices. Here we use the definition in Equation~\eqref{determines}.

\bs \begin{center}
\begin{minipage}{11truecm}
\fbox{{\parbox{1\linewidth}{
{\bf Algorithm Dominant--Vertices.}

\begin{small}
\begin{tt}
\ms Procedure Initial-Solution. \\
Input $\G:=(V,A)$. Output: $U$ and $U'=V\setminus U$.\\
 \begin{itemize}
   \item[] Set $U:=\{u\in V:\ \Id(u)=0\}$, and $U':=\emptyset$.
   \item[] For $i=1$ to $\#V$, $W:=U\cup U'$, $U'=\d W$, End.
   \item[] If $W = V$, then finish, else 
        \begin{itemize} 
         \item[] Set $V':=V \setminus W$.
         \item[] While $W \neq V$, 
             \begin{itemize}
             \item[] Set $U:=U\cup\{u\}$, with $u\in V'$ such that \\
                   $\Id(u)$ minimal, given that $\Od(u)$ is maximal. 
             \item[] For $i=1$ to $\#V$, $W:=U\cup U'$, $U'=\d W$, End.
             \item[] Set $V':=V\setminus W$.
         \end{itemize}
         \item[] End.
      \end{itemize}
   \item[] End.
 \end{itemize}

\ms Procedure Optimization. \\
Input: $U$ and $U'=V\setminus U$. Output: $U$ and $U'=V\setminus U$.\\
   \begin{itemize}
     \item[] Fix an order in $U$, $U:=\{u_1,\ldots,u_{\#U}\}$.
     \item[] For $i=1,2,\ldots,\#U$.
     \begin{itemize}
       \item[] Set $\bU:=U\setminus\{u_i\}$, $\bU':=\d \bU$, and $\bV:=U'\cup\{u_i\}$.
       \item[] For $i=1$ to $\#V$, $W:=\bU\cup \bU'$, $\bU'=\d W$, End.
       \item[] If $\bV\subseteq W$, then $U:=\bU$ and $U'=\bU'$, End.
     \end{itemize}
     \item[] End. 
   \end{itemize}

\ms Procedure Nested. \\
Input: $U$ and $U'=V\setminus U$. Output $U=U_1\subset U_2 \subset U_{d+1}=V$.
   \begin{itemize}
     \item[] Set $U_1:=U$, $W:=U_1$, and $d:=1$. 
     \item[] While $W \neq V$, $U_{d+1}:=U_d\cup \d U_d$, $W:=U_{d+1}$, $d=d+1$, End.
   \end{itemize}

\end{tt}
\end{small}
}}}\end{minipage}\end{center}

\bs Notice that domination is a property completely determined
by the structure of the underlying network $\G:=(V,A)$, and does not depend on 
the particular choice of activation partitions $\P_v$, or interaction function $D$.

\section{Theoretical Examples}~\label{section-thexamples}\

\ms In order to understand the relation between the structure of the underlying 
network and the set of dominant vertices, we present below 
some examples where minimal dominant sets can be explicitly determined. 
For those examples we also know, in a explicit way, the output of the 
algorithm Dominant--Vertices. We can therefore test the accuracy of the
algorithm in this controlled situation. At the end of the section we present
a example to illustrate Proposition~\ref{proposicion1}, concerning  
how the dominance is inherited to completely connected subnetworks.

\ms 
\subsection{Dominant Set of Vertices}

\subsubsection{Regular Trees}\

\ms {\it A $b$--regular tree with $L$ levels} can be constructed as follows. 
From a finite set $B$ 
on  $b$ symbols, we define the vertex set $V:=\cup_{\ell=0}^{L-1} B^\ell$, where 
$B^0\equiv\{\varepsilon\}$ with $\varepsilon$ denoting the empty word. 
The set $B^\ell$ defines the $\ell$--th level of the tree, and the arrows are given by
$A:=\left\{(\omega,\tilde\omega)\in V:\ \tilde\omega\in\{\omega\}\times B \text{ or }
                                         \omega\in\{\tilde\omega\}\times B \right\}$.
In this way, vertices in the $\ell$--th level form arrows only with vertices in the adjacent
levels $B^{\ell\pm 1}$. Because of this, the union of odd levels determine the even ones, and
vice versa, i.~e.
\begin{eqnarray*}
\bigcup_{\ell \text{ odd }} B^{\ell}&=&\d \left( \bigcup_{\ell \text{ even} } B^{\ell} \right)\\
\bigcup_{\ell \text{ even}} B^{\ell}&=&\d \left( \bigcup_{\ell \text{ odd}} B^{\ell}\right).	
\end{eqnarray*}
Hence, any of $V_{\rm even}:=\cup_{\ell \text{ even}} B^{\ell}$ or 
$V_{\rm odd}:=\cup_{\ell \text{ odd}} B^{\ell}$ is a dominant set of vertices. 

\begin{figure}[h!]
\centering
\includegraphics*[width=0.5\textwidth]{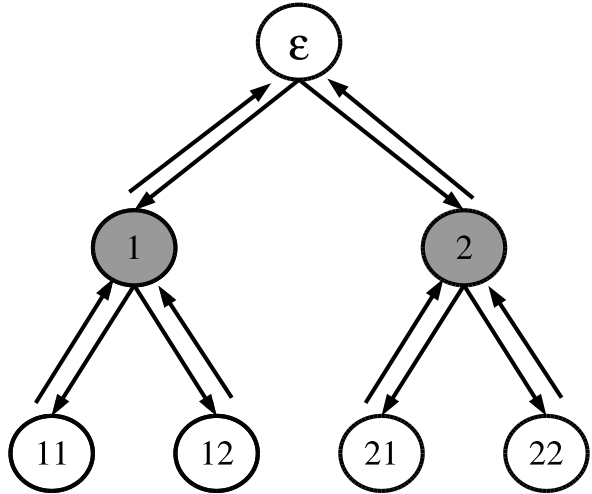}
\caption{The $2$--regular tree with 3 levels. The dominant vertices are colored in grey.}
\label{arb}
\end{figure}

\ms Figure~\ref{arb} shows a $2$--regular tree with 3 levels. In this case we can select 
as dominant set the first level in the tree, which has 2 vertices only. 

\ms If we apply {\tt Procedure Initial-Solution} to the $2$--regular tree with 3 levels, we obtain
$U=\{1,2\}$, and $U'=\{\varepsilon, 11,12,21,22\}$. {\tt Procedure Optimization} 
returns the same output. Finally, {\tt Procedure Nested} gives $d=1$ and 
$U_1=\{1,2\}\subset U_2=V$. In general, for a $b$--regular tree with $L$ level, 
the algorithm Dominant--Vertices returns $d=1$ and $U_1:=V_{\rm odd}\subset U_2=V$. 
It is not hard to show that $V_{\rm odd}$ is the smallest dominant 
set for trees with an odd number of levels. 

\bs 

\subsubsection{Dictatorial Networks}\

\begin{figure}[h!]
\centering
 	\includegraphics*[width=0.5\textwidth]{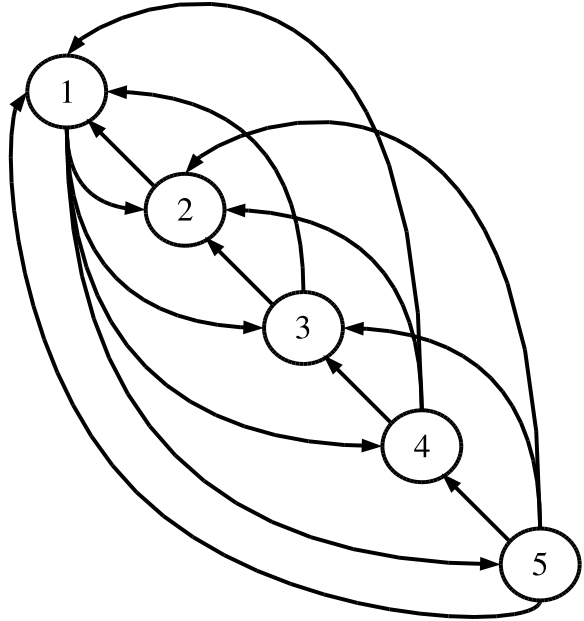}
\caption{Dictatorial network on $5$ vertices.}
\label{gi}
\end{figure}

\ms In this paragraph we present a collection of completely connected networks with a single
dominant vertex. For $n\geq 2$, consider the networks with vertex set
$V:=\{1,2,\ldots,n\}$, and arrows defined by the adjacency matrix 
\[
M_A:=\left(\begin{matrix}
0&1&1&\dots&1&1&1\\
1&0&0&\dots&0&0&0\\
1&1&0&\dots&0&0&0\\
1&1&1&\dots&0&0&0\\
\vdots&&&&&&\vdots\\
1&1&1&\dots&1&0&0\\
1&1&1&\dots&1&1&0\\
\end{matrix}\right).
\]
The resulting digraph, which we call {\it dictatorial network on $n$ vertices}, was a total of
$(n-1)(n+2)/2$ arrows. Figure~\ref{gi} shows the dictatorial network on $5$ vertices, 
which allows us to understand why vertex number $1$ is dominant in all dictatorial networks. 
Here, according to Equation~\eqref{determines},
vertex 1 determines vertex 5, then $\{1,5\}$ determines vertex number $4$. The set $\{1,4,5\}$ 
determines vertex $3$, and finally $\{1,3,4,5\}$ determines vertex number $2$.  

\ms If we apply {\tt Procedure Initial-Solution} to the dictatorial networks in 
Figure~\ref{gi}, we obtain $U=\{1,2,3,4,5\}$ and $U'=\emptyset$. {\tt Procedure Optimization}
returns $U=\{1\}$ and $U'=\{2,3,4,5\}$. Finally, {\tt Procedure Nested} gives $d=4$ and 
$U_1:=\{1\}\subset U_2:=\{1,5\}  \subset U_3:=\{1,4,5\} \subset U_4:=\{1,3,4,5\}\subset 
U_5:=\{1,2,3,4,5\}=V$. In general, since the dictatorial network on $n$ vertices is such that
\begin{eqnarray*}
    \d \{1\}  &    = &\{n\} \\
    \d \{1,n\}&    = &\{n-1\}\\
\d \{1,n-1,n\}&    = &\{n-2\}\\
               &\vdots&  \\
\d \{1,3,\ldots,n\}&=&\{2\},
\end{eqnarray*}           
then algorithm Dominant--Vertices returns $d=n-1$ and
$U_1=\{1\}\subset U_2=\{1,n\}  \subset U_3=\{1,n-1,n\} \dots \subset U_{n}=V$. 
Of course, the dominant set that is produced by the algorithm is of minimal cardinality.

\bs \subsubsection{Democratic Networks}\

\ms In contrast to the previous family of examples, here we present a collection of 
networks on $2n$ vertices, whose dominant sets have cardinality of at least $n$. 
They are bi--directional circuits on an even number of vertices, which we define as follows.
The vertex set is $V:=\{0,2,\ldots,2n-1\}$, and arrows
$A:=\{(u,v)\in V\times V: \ u-v\equiv 1 (\mod 2n) \ \text{ or } u-v \equiv 2n-1 (\mod 2n)\}$.
The corresponding adjacency matrix has the form
\[
M_A:=\left(\begin{matrix}
0&1&0&0&\dots&0&0&1\\
1&0&1&0&\dots&0&0&0\\
0&1&0&1&\dots&0&0&0\\
0&0&1&0&\dots&0&0&0\\
\vdots&&&&&&&\vdots\\
0&0&0&0&\dots&1&0&1\\
1&0&0&0&\dots&0&1&0\\
\end{matrix}\right).
\]
We will refer to the network just defined as 
{\it the democratic network on $2n$ vertices}. 

\begin{figure}[h!]
\centering
 	\includegraphics*[width=0.5\textwidth]{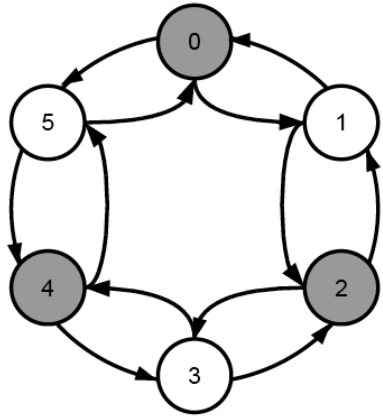}
\caption{Democratic network on $6$ vertices.}
\label{dem2}
\end{figure}

\ms Figure~\ref{dem2} shows the democratic network on $6$ vertices, 
the set $\{0,3,4\}$ determines $\{1,3,5\}$, and vice versa.

\ms Applying {\tt Procedure Initial-Solution} to the democratic networks in 
Figure~\ref{dem2}, we obtain $U=\{0,1,2,3,4\}$ and $U'=\emptyset$. {\tt Procedure Optimization}
returns $U=\{0,2,4\}$ and $U'=\{1,3,5\}$. Finally, {\tt Procedure Nested} gives $d=1$ and 
$U_1:=\{0,2,4\}\subset U_2:=\{0,1,2,3,4,5\}=V$. In general, algorithm Dominant--Vertices returns 
$d=1$ and $U_1=\{0,2,4,\ldots, 2n-2\}\subset U_2=\{0,1,2,3,\ldots, 2n-1\}=V$. This is because
\begin{eqnarray*}
    \d \{0,2,4,6,\ldots,2n-2\}&    = &\{1,3,5,\ldots,2n-1\}\\
    \d \{1,3,5\ldots,2n-1\}  &    = &\{0,2,4,6,\ldots, 2n-2\}.
\end{eqnarray*}           
Notice that the dominant set that is produced by the algorithm is of minimal cardinality.

\ms 
In the three examples examined so far, the dominant set $U$ determined by the Algorithm 
Dominant--Vertices is of minimal cardinality, and the complement $U'=V\setminus U$ 
is dominant as well, but this is not the general case. Furthermore, even when 
$U'=V\setminus U$ is dominant, the number of steps needed to determine $U$ from $U'$ 
can be different from the number $d$ given by the algorithm. This is in particular 
the case of the dictatorial networks.

\ms\subsection{Inhering Dominance}\

\begin{center}
\begin{figure}[h!]~\label{lem1}
\begin{tabular}{ccc}
\includegraphics*[width=0.33\textwidth]{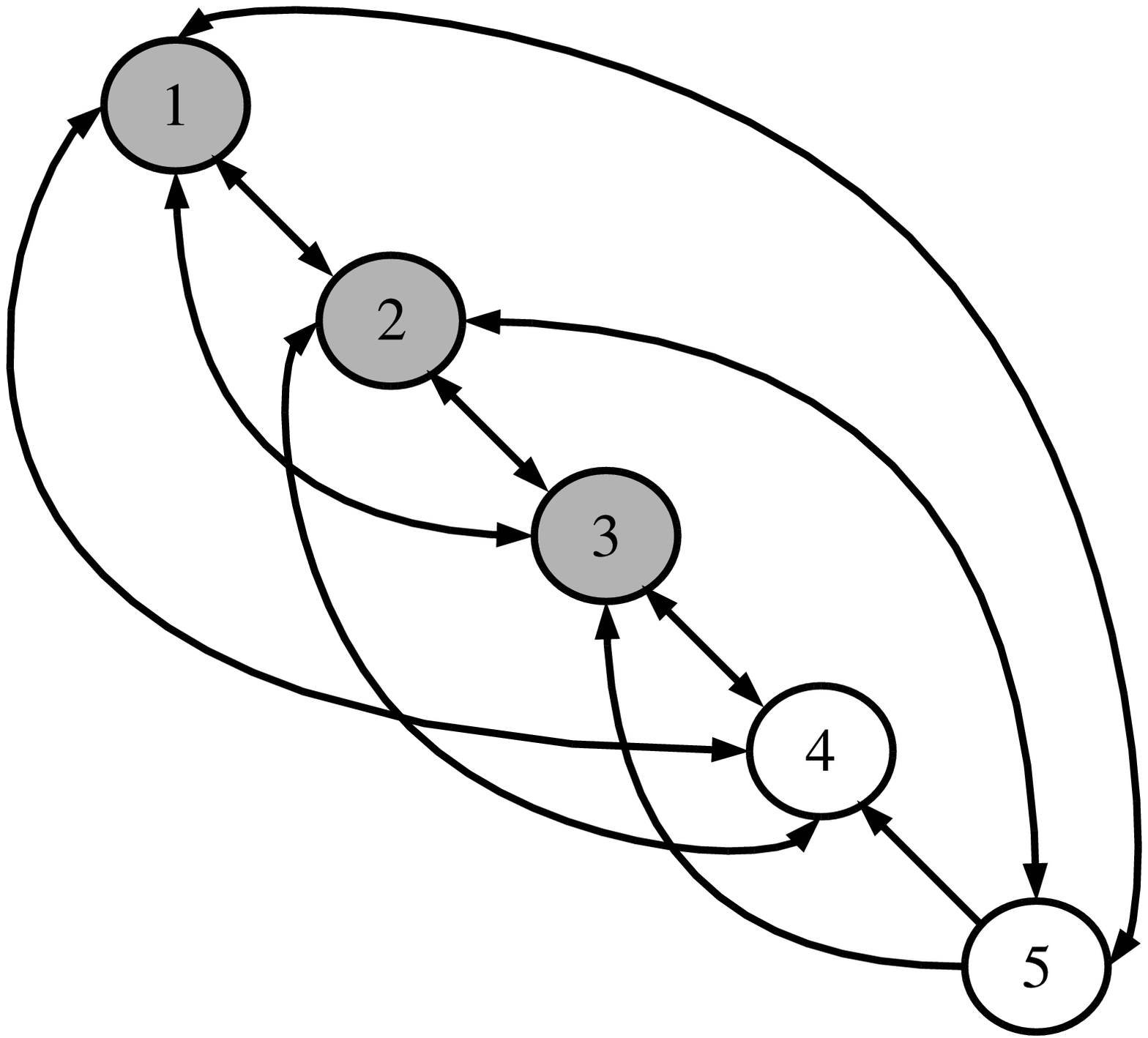}
&\includegraphics*[width=0.33\textwidth]{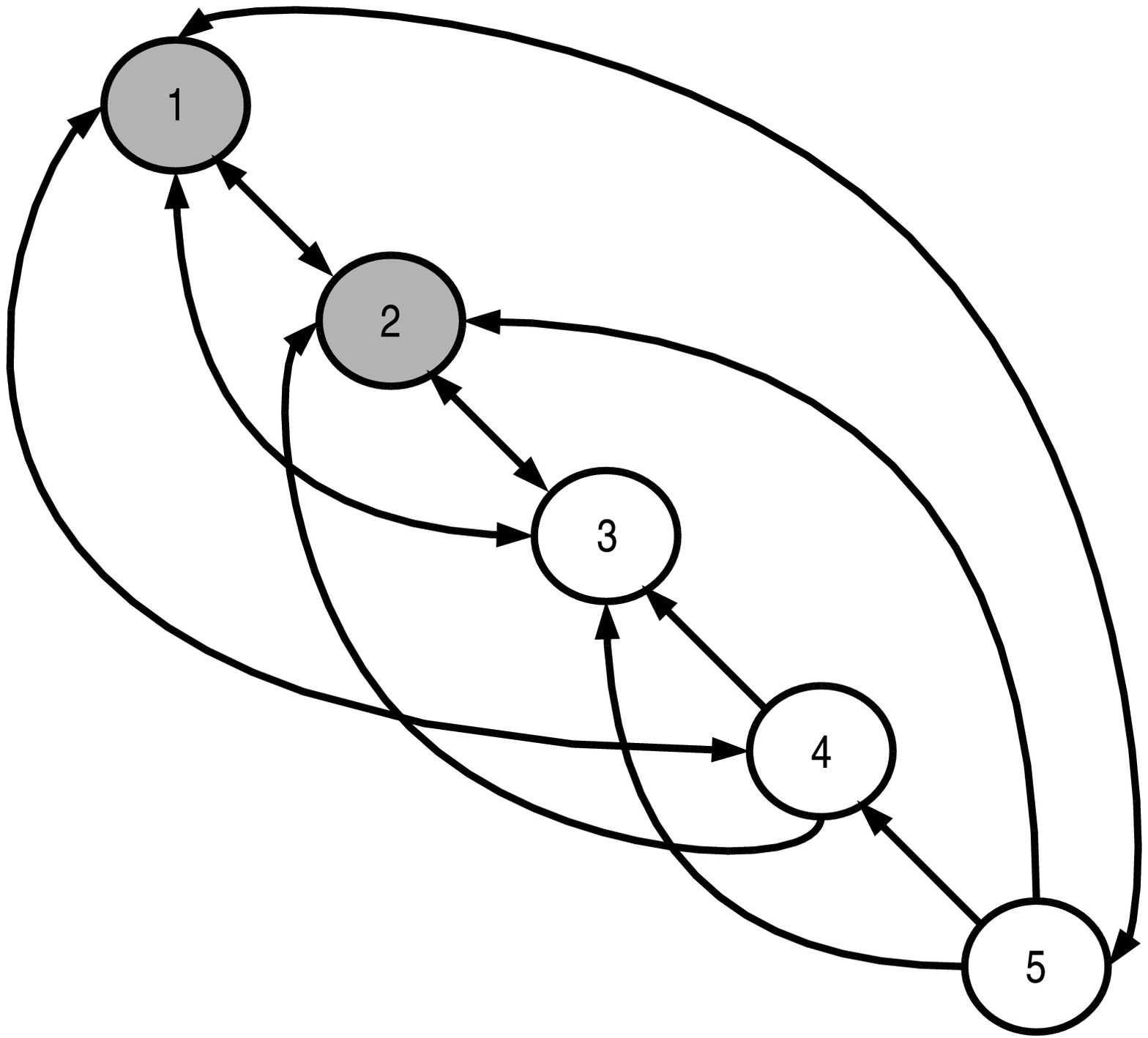}
&\includegraphics*[width=0.33\textwidth]{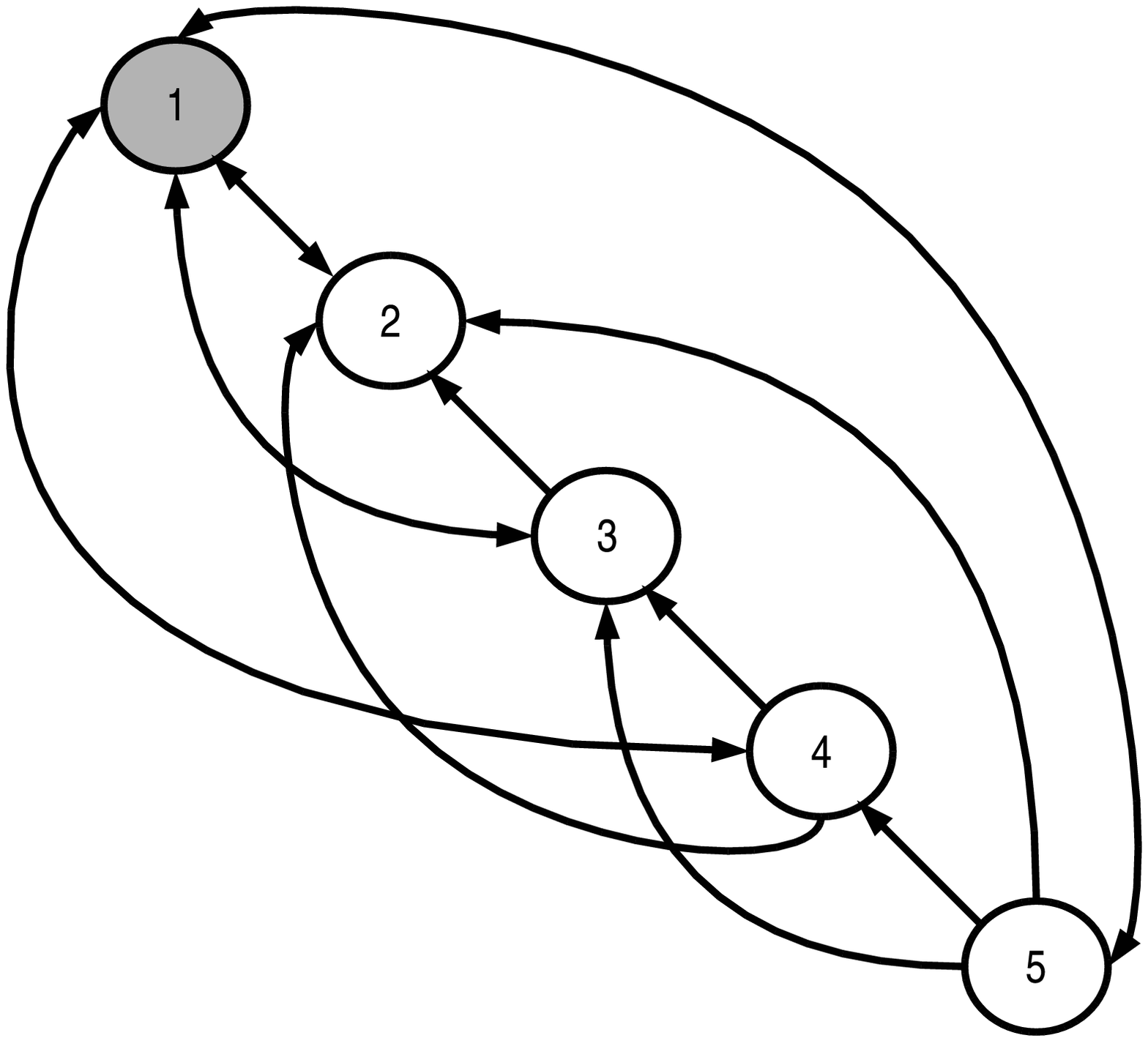}
\end{tabular}
\caption{Supernetworks of the dictatorial network on $5$ vertices. Their
respective dominant vertices are colored in gray.}
\end{figure} 
\end{center}

\ms To illustrate how dominance is inherited to subnetworks, we will consider three supergraphs
of the dictatorial network in Figure~\ref{gi}, which we present in Figure~\ref{lem1}.
Starting with the network at left, we obtained the other two by eliminating an increasing number
of arrows, keeping a strongly connected network each time.
For the graph at left, $\{1,2,3\}$ is the dominant set of vertices given by the Algorithm 
Dominant--Vertices. Applying the same algorithm to the network at the center of the figure,
we obtain $\{1,2\}$ as the dominant set. Finally,  for the dictatorial network on $5$ vertices,
vertex number $1$ is the dominant vertex the algorithm returns.  

\section{Numerical Examples}~\label{section-nuexamples}\

\ms In this section we consider two families of random networks, 
and we determine a dominant set of vertices by means of the Algorithm 
Dominant--Vertices. We will characterize the structural complexity
of the network through the expected size of the dominant set $U$ and 
the number of steps $d$ needed to determine $V$ from $U$. 

\subsection{Erdo\"s--R\'enyi networks}\

\ms An Erdo\"s--R\'enyi digraph is constructed as follows. Fix a set of 
vertices $V$ and $p\in (0,1]$. The arrow set $A\subseteq V\times V$ is such 
that for each $u,v\in V$, $\pp\{(u,v)\in A\}=p$. According to this, a given 
arrow $(u,v)$ belongs to the digraph with probability $p$, and the 
inclusion of different vertices are independent and identically distributed 
random variables (for more details on random graphs and their dynamical 
construction see~\cite{D07} for instance). 

\begin{figure}[h!]
\begin{center}\begin{tabular}{c}
\includegraphics*[width=0.8\textwidth]{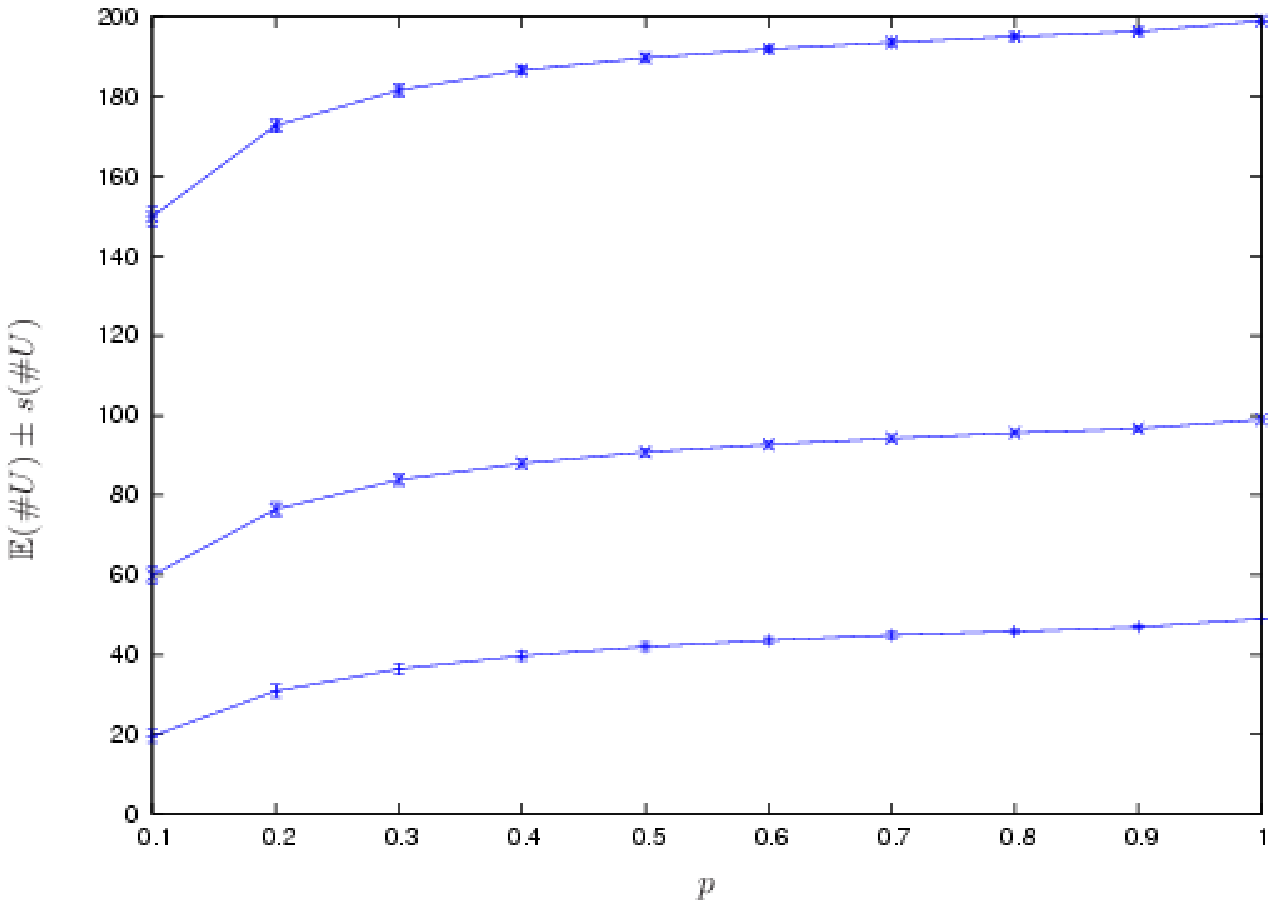} \\ \includegraphics*[width=0.8\textwidth]{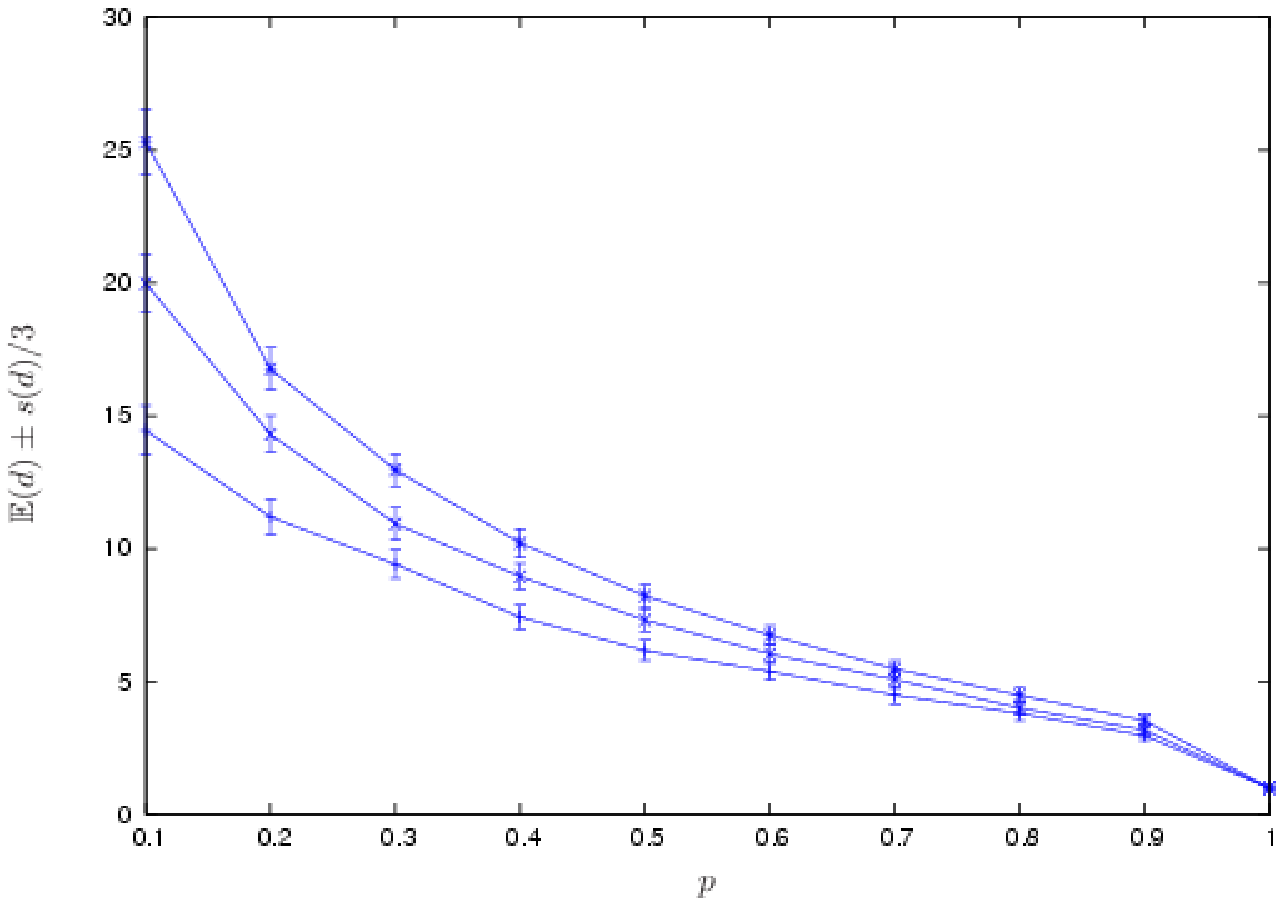}
\end{tabular}\end{center}
\caption{On top, the expected number of dominant vertices $\ee(\#U)$,
with error bars corresponding to the fluctuation $s(\#U):=\sqrt{\ee(\#U-\ee(\#U))^2}$, 
both as a function of $p$.
On bottom, the mean depth $\ee(d)$, with error bars proportional to the 
corresponding standard deviation $s(d):=\sqrt{\ee(d-\ee(d))^2}$, 
both as a function of the connection probability $p$.
In both frames we compare the three scenarios $\#V=50,100$ and $200$,  and  
their corresponding curves appeared ordered from bottom to top.}
\label{erdos-fig}
\end{figure}

\ms For $p=0.1,0.2,\ldots,1$ and $\#V=50,100,200$, we have generated 
100 Erdo\"s--R\'enyi digraphs, and for each one of those 3000 networks, we 
have computed a dominant set of vertices $U$ and a depth $d$, by using 
Algorithm Dominant--Vertices. 
On top of Figure~\ref{erdos-fig} we show the expected number of dominant 
vertices $\ee(\#U)$ the Algorithm Dominant--Vertices returns, and 
in error bars the size of the fluctuations $s(\#U):=\sqrt{\ee(\#U-\ee(\#U))^2}$, 
both as a function of $p$, and for the three scenarios $\#V=50,100,200$. In the 
lower frame of the same figure we plot the expected number of steps $\ee(d)$ 
given by the same algorithm, with error bars proportional to the corresponding 
standard deviation $s(d):=\sqrt{\ee(d-\ee(d))^2}$, both as a function of the 
connection probability $p$.

\begin{figure}[h!]
\centering
\includegraphics*[width=0.6\textwidth]{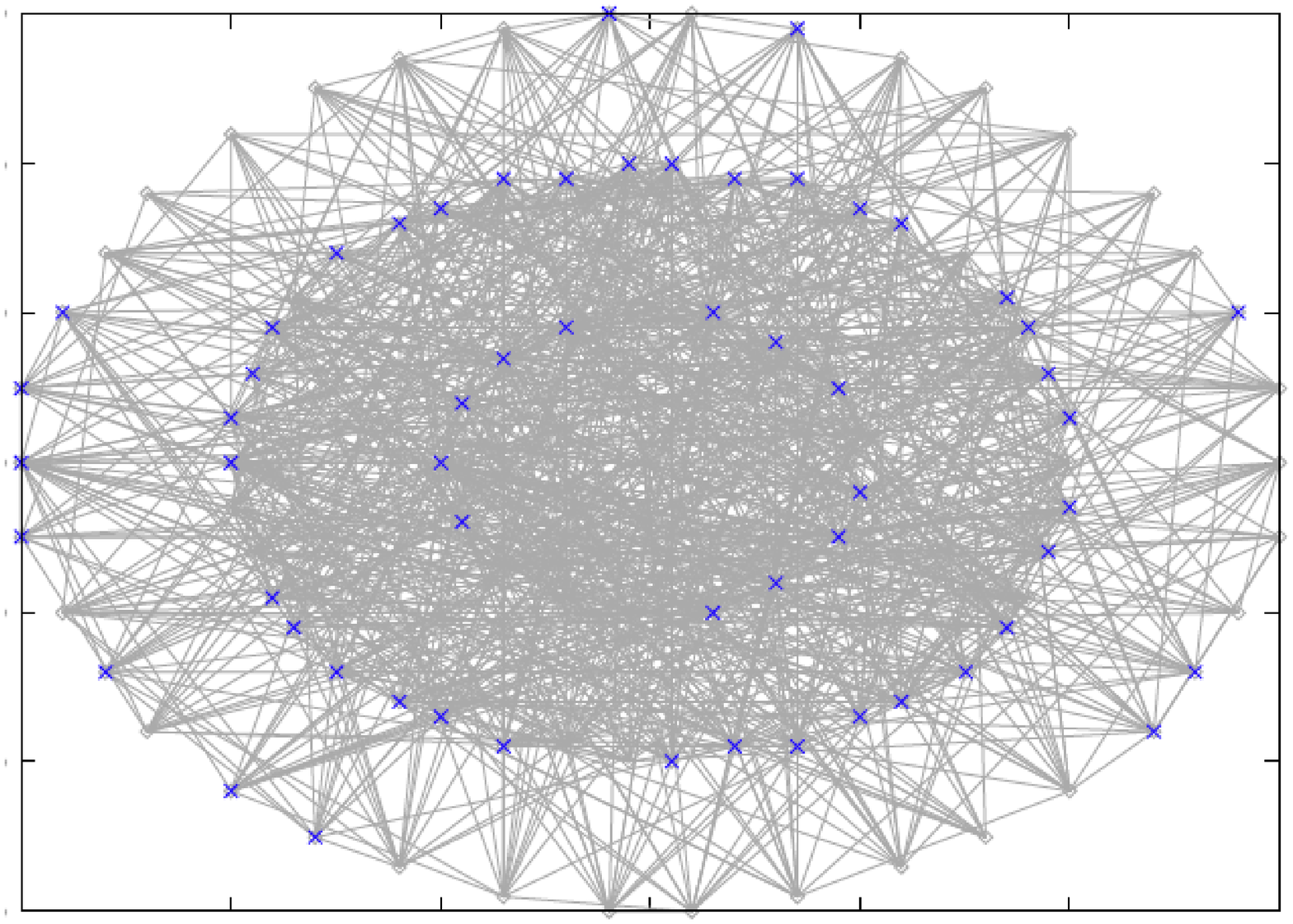}
\caption{Realization of an Erdo\"s--R\'enyi Network on 100 vertices. The dominant vertices
determined by the Algorithm Dominant--Vertices are marked with a $\times$.}
\label{erdos-exo}
\end{figure}

\ms In Figure~\ref{erdos-exo} we show a realization of an Erdo\"s--R\'enyi network on $100$ vertices, 
with probability of connection $p=0.1$. For this example, the Algorithm Dominant--Vertices
retunrs $\#U=58$ and $d=13$.

\ms For the completely connected network on $N$ vertices, any dominant set
has cardinality $N-1$ and depth $d=1$. Our numerical results show how the
Erdo\"s--R\'enyi networks approaches this situation as the probability of 
connection goes to 1. More specifically, as $p\to 1$, the size of the dominant
set $\ee(\#U)\to N$, while $\ee(d)\to 1$.

\subsection{Barab\'asi--Albert Networks}\

\ms We implemented a dynamical construction of scale--free graphs similar to the one
proposed by Barab\'asi and Albert. We start, at $n=0$, with a kernel graph $\G_0$, 
which is the complete simple graph on $m_0$ vertices. Then, for each $n\geq 1$, we add a 
new vertex $v_{n+1}$ to the preceding graph $\G_n:=(E_n,V_n)$. This new vertex form new edges with 
randomly chosen vertices in $V_n$. A given vertex $v\in V_n$ has probability proportional 
to its degree to be selected, and we have
\[
\{v_{n+1},v\}\in E_n \text{ with probability } 
p_n(v):=\frac{\#\{ u\in V_n: \ \{u,v\}\in E_n\}}{\#E_n}.
\] 
The graph at the $(n+1)$--th step is $\G_{n+1}$, with $V_{n+1}:=V_n\cup \{v_{n+1}\}$,
and an enlarged edge set $E_{n+1}$. This random iteration continues until a predetermined 
number of vertices is obtained. In our dynamical construction it is possible to add several 
edge at each iteration, in opposition to the traditional scheme where only one edge is added
at each time step. We finally assign directions to the edges in $E_{N-m_0}$,
by choosing one of the two possible directions with probability $1/4$,
and both directions with probability $1/2$ (for a rigorous study of scale--free and related
random graphs, see~\cite{D07}).

\begin{figure}[h!]
\begin{center}\begin{tabular}{c}
\includegraphics*[width=0.8\textwidth]{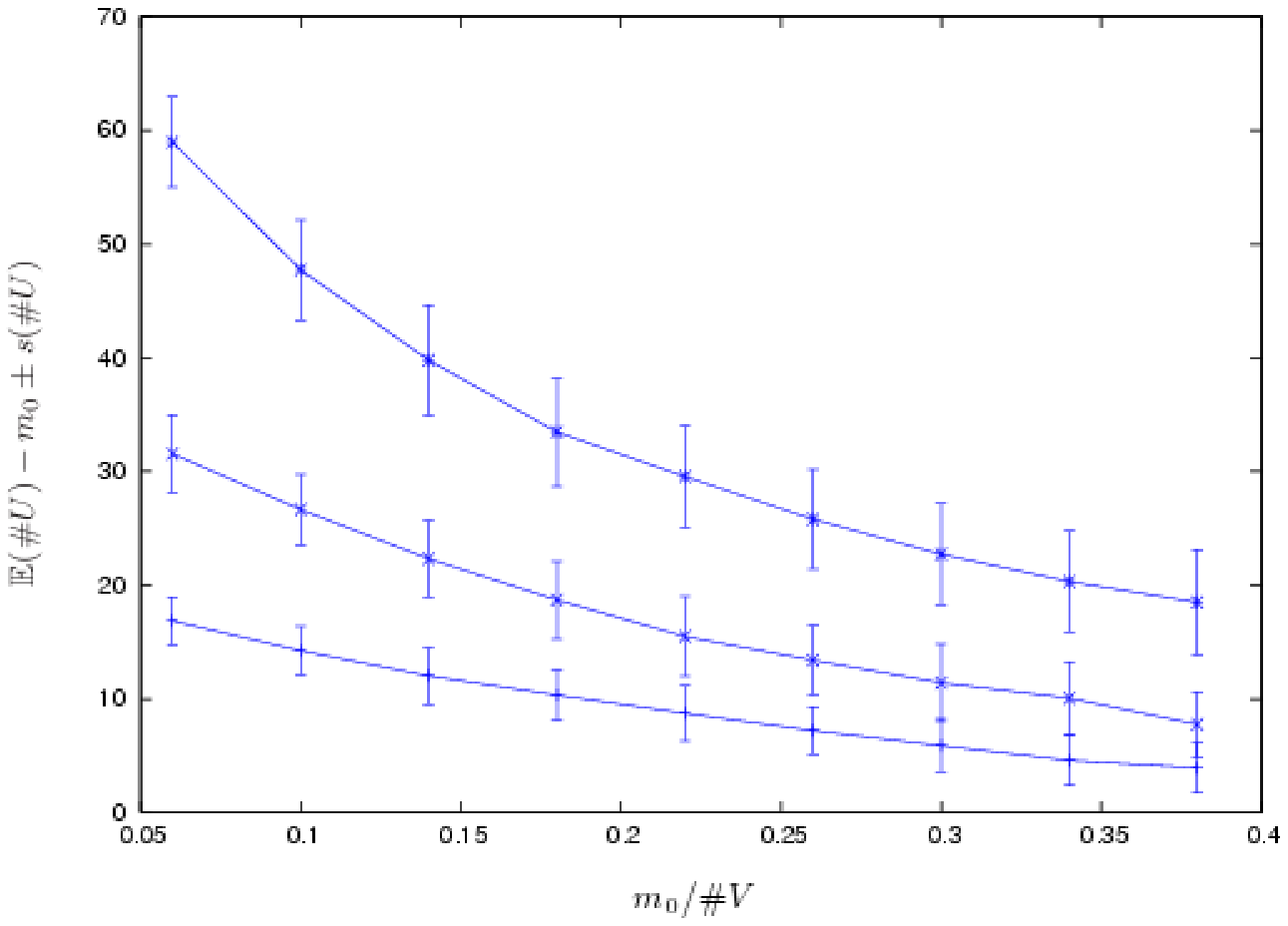} \\ \includegraphics*[width=0.8\textwidth]{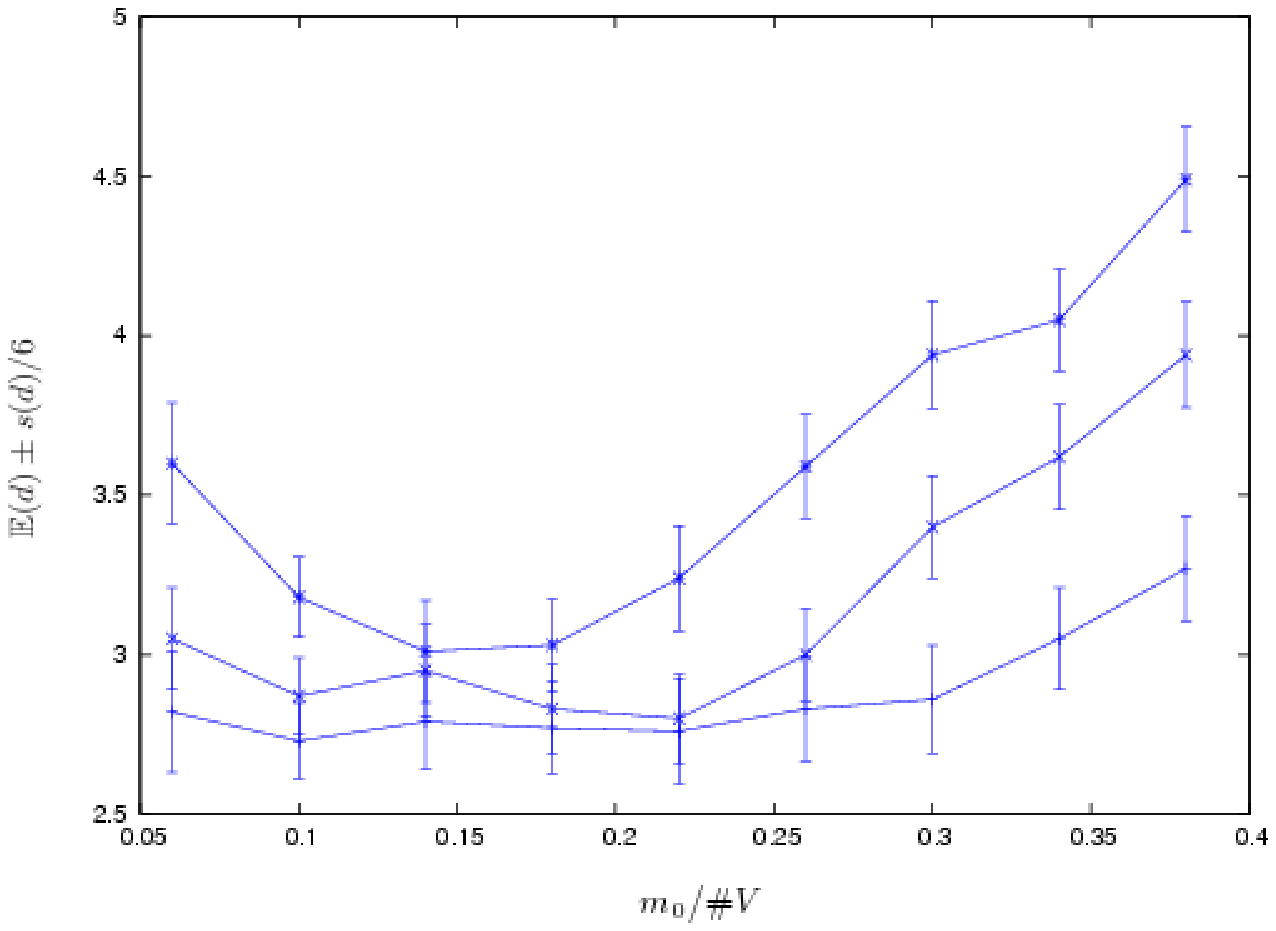}
\end{tabular}\end{center}
\caption{In top, the behavior of $\ee(\#U)-m_0$ as a function of the relative 
size of the kernel $m_0/\#V$. The fluctuations $s(\#U):=\sqrt{\ee(\#U-\ee(\#U))^2}$ 
are indicated by the error bars. At bottom, the mean depth $\ee(d)$, and error bars proportional to the corresponding standard deviation $s(\#d):=\sqrt{\ee(d-\ee(d))^2}$, 
both as functions of the relative size of the kernel.
In both frames we compare the three scenarios $\#V=50,100$ and $200$,  and  
their corresponding curves appeared ordered from bottom to top.}
\label{barabasi-fig}
\end{figure}

\ms For $\#V=50,100,200$, and $m_0=0.06\times\#V,0.10\times\#V,0.14\times\#V,\ldots,0.38\times\#V$, 
we have generated 100 Barab\'asi--Albert networks, and for each one of those 2700 digraphs, 
we have computed $U$ and $d$, by using Algorithm Dominant--Vertices. 
Figure~\ref{barabasi-fig} is the analogous to Figure~\ref{erdos-fig} above. It shows in the upper
frame the behavior
of $\ee(\#U)-m_0$, the expected number of dominant vertices exceeding the kernel of size, 
as a function of the relative size of the kernel $m_0/\#V$. 
In error bars we plot the size of the corresponding fluctuations $s(\#U):=\sqrt{\ee(\#U-\ee(\#U))^2}$. 
The frame at bottom shows the mean depth $\ee(d)$ given by 
the same algorithm, and error bars proportional 
to the corresponding standard 
deviation $s(\#d):=\sqrt{\ee(d-\ee(d))^2}$, also as a function of 
the relative size of the kernel.
In both frames we compare the three scenarios $\#V=50,100$ and $200$.

\begin{figure}[h!]
\centering
\includegraphics*[width=0.6\textwidth]{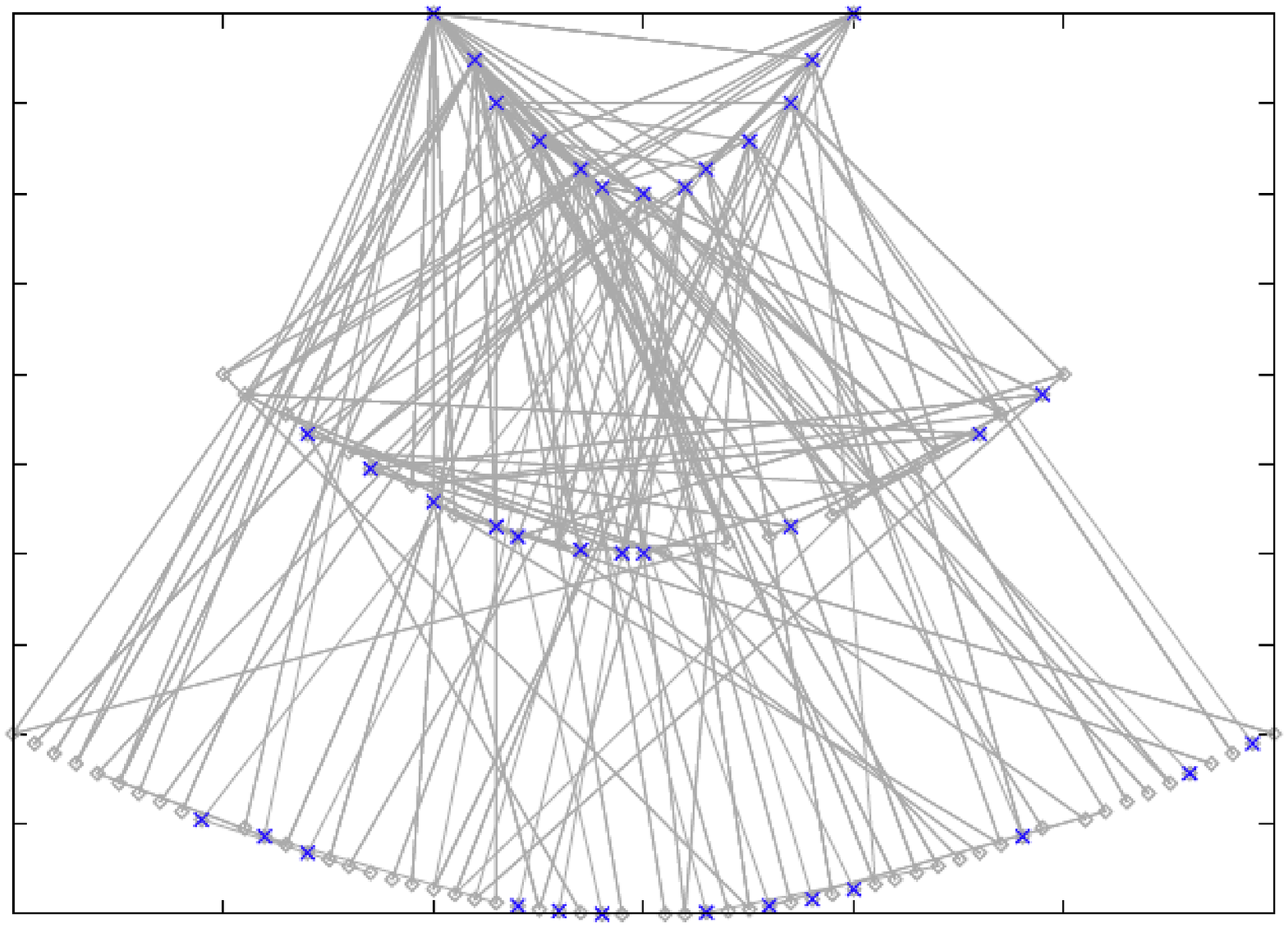}
\caption{Realization of a Barab\'asi--Albert Network on 100 vertices. The dominant vertices
determined by the Algorithm Dominant--Vertices are marked with a $\times$.}
\label{barabasi-exo}
\end{figure}

\ms In Figure~\ref{barabasi-exo} we show a realization of a 
Barab\'asi--Albert network on $100$ 
vertices, with kernel size $m_0=6$. For this example, the Algorithm 
Dominant--Vertices retunrs $\#U=37$ and $d=3$.

\ms We observed that the size of the dominant set for a Barab\'asi--Albert network
increases as the relative size of the kernel grows, in a similar way as in 
the Erdo\"s--R\'enyi case. Nevertheless, there is an important difference.
Since the kernel is build from a complete graph, the size of the dominant set is 
expected to be at least as large as the kernel. We observe that 
number extra vertices needed to complete the dominant set decreases with 
the size of the kernel. At the same time, the depth of the dominant set 
follows a non--monotonic behavior.

\section{Final comments}~\label{section-final}\

\ms Our definition of dominant vertices depends only on the structure 
of the underlying network, and for orbits (or forced trajectories) uniform 
separated from the discontinuity set, those vertices capture the 
dynamical state of the whole network. In the general case, for orbits 
accumulating at the discontinuity set, another definition of domination should be
considered. Nevertheless, since uniform separation from the discontinuity set 
is typical, our algorithm determines observation (or control) nodes, 
in all but a negligible proportion of the cases.  

\ms One could interpret our main result as the existence of a effective
network, smaller than the underlying network, which could be constructed 
from the dominant set of vertices. The dynamics on the 
effective network would be equivalent to the original one. Though such a
network reduction can be attempted, specially in the case of $a=0$ 
(instantaneous degradation), the resulting dynamical system is not necessarily 
simpler that the original one.  
 
\ms The algorithm we propose is non--deterministic,  and several alternative 
formulations can be considered. For instance, in the procedure {\tt Initial-Solution}
one chooses vertices with minimal input degree, amongst those of maximal output degree.
Alternatively, one could choose vertices maximizing the output degree, amongst 
those of minimal input degree. We have considered this alternative, and we have 
found that in general it produces larger dominant sets. In spite of this, we expect 
that this algorithm can be improved, but that was not our aim in this work. 
Instead, we have tested the efficiency of the algorithm on examples where the 
dominant sets of vertices can be explicitly determined, and we have found optimal
or almost optimal results. 

\ms Finally, the properties of dominant sets of vertices can be used to characterize
the complexity of a given network. We have done this for the Erdo\"s--R\'enyi and
for the Barab\'asi--Albert ensembles of random networks. This topological characterization
has to be contrasted with a dynamical one, and this is what we intend to do in
a future work.

\end{document}